\documentclass[epj]{webofc}
\usepackage[varg]{txfonts}  
\usepackage{color}
\woctitle{QCD@Work 2014}

\begin{document}

\title{Multi-strange baryon production in \mbox{pp}, \mbox{p--Pb} and \mbox{Pb--Pb} collisions measured with ALICE}
\author{Domenico Colella\inst{1}\fnsep\thanks{\email{domenico.colella@cern.ch}} (for the ALICE Collaboration)}
\institute{Dipartimento Interateneo di Fisica ''M. Merlin'' and Sezione INFN, Via Orabona 4, 70126 Bari, Italy}

\abstract{The production of $\Xi^{-}$ and $\Omega^{-}$ baryons and their anti-particles in pp, \mbox{p--Pb} and \mbox{Pb--Pb} collisions has been measured by the ALICE Collaboration. These hyperons are reconstructed via the detection of their charged weak-decay products, which are identified through their measured ionisation losses and momenta in the ALICE Time Projection Chamber. Comparing the production yields in \mbox{Pb--Pb} and \mbox{pp} collisions, a strangeness enhancement has been measured and found to increase with the centrality of the collision and with the strangeness content of the baryon; moreover,  in the comparison with similar measurements at lower energies, it decreases as the centre-of-mass energy increases, following the trend already observed moving from SPS to RHIC. Recent measurement of cascade and $\Omega$ in \mbox{p--Pb} interactions are compared with results in \mbox{Pb--Pb} and \mbox{pp} collisions and with predictions from thermal models, based on a grand canonical approach.
The nuclear modification factors for the charged $\Xi$ and $\Omega$, compared to the ones for the lighter particles, are also presented. 
}

\maketitle

\section{Introduction}
\label{sec-0}
Measuring strange and multi-strange particle production in relativistic heavy-ion interactions is a unique tool to investigate the properties of the hot and dense matter created in the collision, as there is no net strangeness content in the initially colliding nuclei. Moreover, the high energy and luminosity available allow for a relative abundant production of such particles. 

An enhanced production of strange particles in \mbox{A--A} compared to pp interactions was one of the earliest proposed signatures of the formation of a deconfined Quark-Gluon Plasma (QGP) \cite{Rafelski}. In \mbox{pp} collisions, the production of strange particles is limited by the local canonical strangeness conservation. This is not the case in central heavy-ion collisions, where strange quarks are expected to be produced more abundantly, resulting in an enhanced production of strange baryons at the end of the  hadronisation stage. 
Strangeness enhancement in heavy-ion collisions with respect to \mbox{pp} (\mbox{p--Be}) has been observed in \mbox{Pb--Pb} collisions at the NA57 experiment \cite{strangEnhancNA57}, at STAR \cite{strangEnhancSTAR} in \mbox{Au--Au} collisions, and indeed confirmed by ALICE in central \mbox{2.76 TeV} \mbox{Pb--Pb} collisions \cite{strangEnhancALICE}.

One of the most striking differences between the \mbox{A--A} and \mbox{pp} collisions first observed at RHIC was the suppression of the hadron production at high transverse momentum ($p_{\rm T}$) in central \mbox{Au--Au} collisions at $\sqrt{s_{\rm NN}}$ = \mbox{200 GeV} when compared to expectations from an incoherent superposition of nucleon-nucleon interactions \cite{RHICRaa}. The effect, confirmed by the charged-particle nuclear modification factor ($R_{\rm AA}$) as a function of $p_{\rm T}$ measured in \mbox{Pb--Pb} collisions at the LHC \cite{ALICERaa}, is interpreted as the result of the energy loss suffered by partons when traversing the hot and dense matter created in an ultra-relativistic heavy-ion collision. The larger suppression observed at LHC compared to that at RHIC can be explained by the higher density of the medium created in collisions at higher centre-of-mass energy.

\section{Multi-strange decay reconstruction with the ALICE detector}
\label{sec-1}
The ALICE detector was designed to study heavy-ion physics at the LHC. It consists of a central barrel with a large solenoid providing a \mbox{0.5 T} field for tracking and particle identification and forward detectors for triggering and centrality selection. Tracking and vertexing are performed using the Inner Tracking System (ITS), consisting of six layers of silicon detectors, and the Time Projection Chamber (TPC). The two innermost layers of the ITS and the V0 detector (scintillation hodoscopes placed on either side of the interaction region) are used for triggering. The V0 also provides the centrality determination in \mbox{Pb--Pb} collisions as well as the multiplicity class selection in \mbox{p--Pb} collisions. A complete description of the ALICE sub-detectors can be found in \cite{JINST}.

Multi-strange baryons are reconstructed through their weak decay topologies, namely \mbox{$\Xi^{-} \rightarrow \pi^{-} + \Lambda$}, \mbox{$\Omega^{-} \rightarrow \textrm{K}^{-} + \Lambda$} (with \mbox{$\Lambda \rightarrow\pi^{-}  + \textrm{p}$}), and the corresponding charge conjugate decays for the anti-particles. The branching ratios are 63.9\% and 43.3\%, for the $\Xi$ and the $\Omega$, respectively. The $\Xi$ and the $\Omega$ candidates are found by combining reconstructed charged tracks: cuts on geometry and kinematics are applied to first select the $\Lambda$ candidate and then to match it with all the remaining secondary tracks (bachelor candidates). In addition, particle identification is performed by selecting on specific energy loss in the TPC for the three daughter tracks.

The signal in $p_{\rm T}$ intervals is obtained by fitting the invariant mass peak ($\pm3\sigma$) with a sum of a Gaussian and a polynomial.
The background is sampled in two regions on both sides of the peak and fitted with a polynomial of first or second degree (depending on the colliding system and the $p_{\rm T}$ interval). More details can be found in \cite{strangEnhancALICE}.

\section{Strangeness enhancements}
\label{sec-2}
\begin{figure}
\centering
\includegraphics[width=20pc,clip]{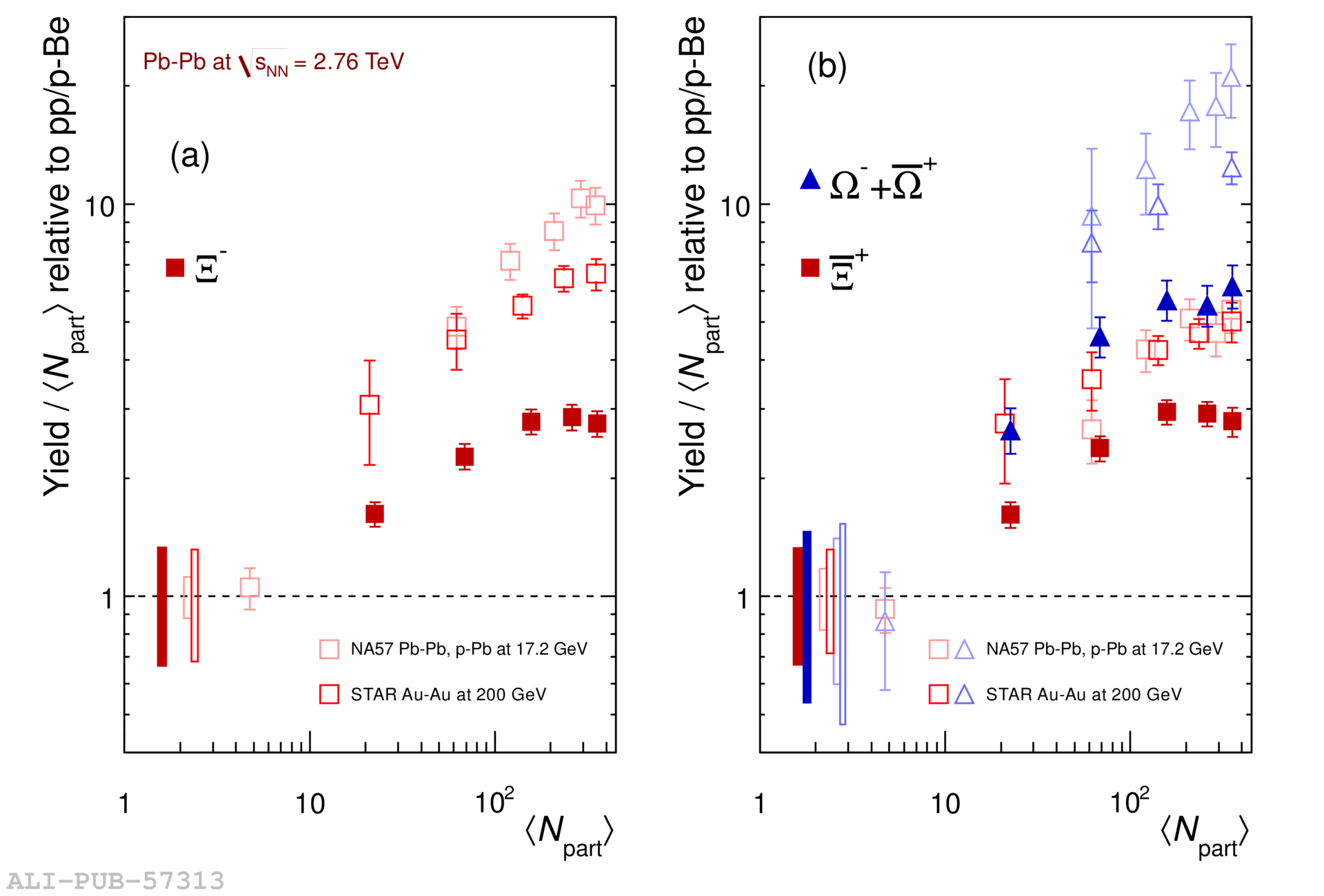}
\includegraphics[width=10pc,clip]{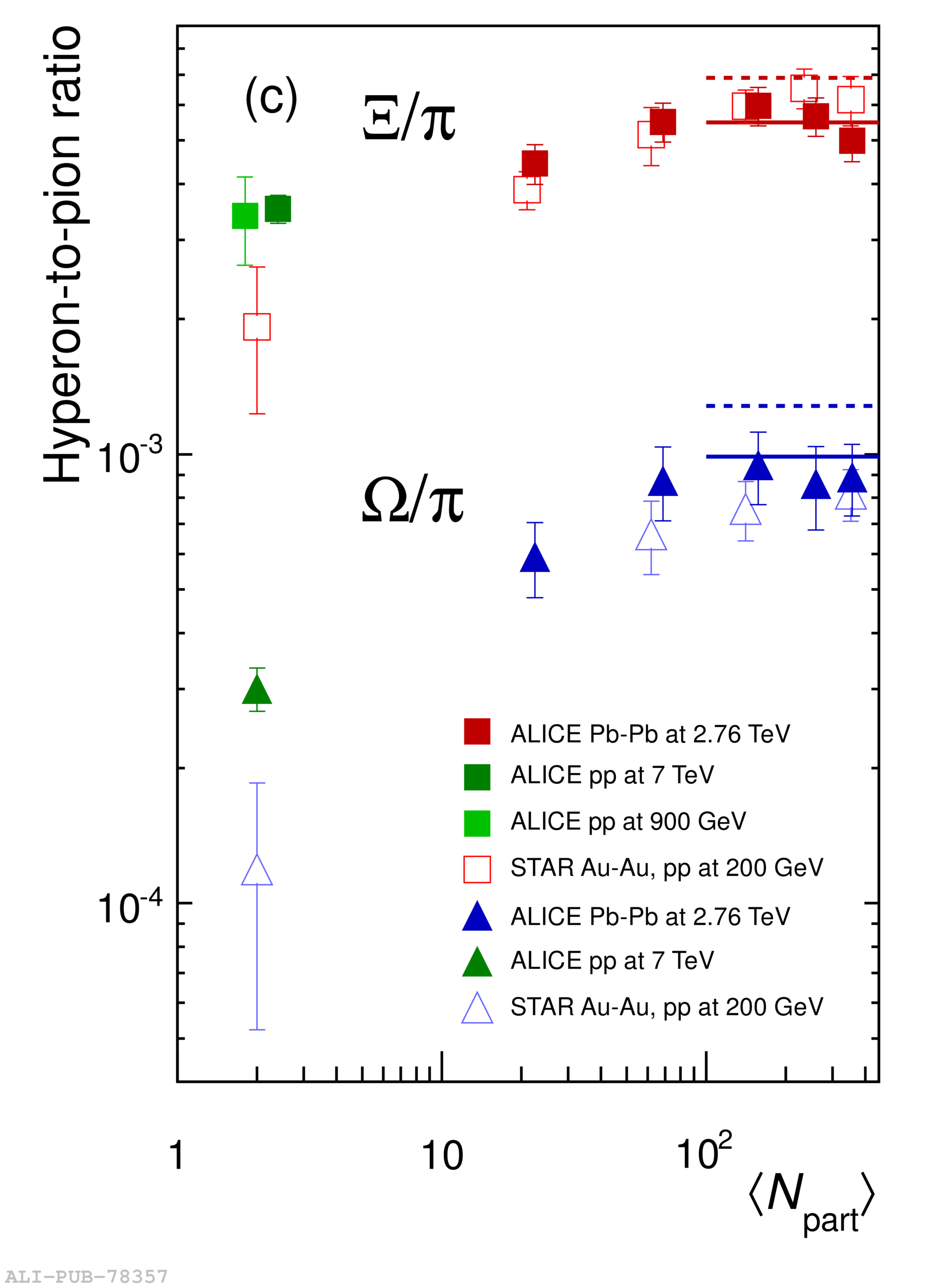}
\caption{(a), (b) Enhancements in $|y|$ $<$ 0.5 as a function of the mean number of participants $N_{\rm part}$ measured by ALICE and compared to SPS and RHIC data. The bars on the dotted line indicate the systematic uncertainties on the pp reference. (c) Hyperon-to-pion ratios as a function of $\langle{N_{\rm part}\rangle}$, for \mbox{A--A} and pp collisions at LHC and RHIC energies. The lines mark the thermal model predictions from \cite{GSImodel} (full line, \mbox{T = 164 MeV}) and \cite{THERMUS} (dashed line, \mbox{T = 170 MeV}).}
\label{fig-1}  
\end{figure}

The strangeness enhancements, the ratio between the yields in \mbox{Pb--Pb} collisions and those in \mbox{pp} interactions at the same energy, both normalized to the mean number of participants ($\langle{N_{\rm part}\rangle}$), are shown in Figure \ref{fig-1}a and \ref{fig-1}b. The \mbox{pp} reference values were obtained by interpolating ALICE and STAR data at different energies and using the excitation function from PYTHIA Perugia-2011 tune \cite{strangEnhancALICE}. This reference agrees with the preliminary results of the yields in \mbox{pp} collisions at the reference energy, \mbox{2.76 TeV}. The enhancement as a function of $\langle{N_{\rm part}\rangle}$ increases with centrality and with the strangeness content of the particle as already observed at lower energies, and decreases as the centre-of-mass energy increases, continuing the trend established between SPS and RHIC energies.

An alternative way to look at the strangeness enhancement, in order to factor out the general increase in multiplicity with \textit{N}$_{\rm part}$ from the actual increase in strange particle production is to study the hyperon-to-pion ratios \mbox{$\Xi/\pi\equiv{(\Xi^{-}+\overline{\Xi}^{+})/(\pi^{-}+\pi^{+})}$} and \mbox{$\Omega/\pi\equiv{(\Omega^{-}+\overline{\Omega}^{+})/(\pi^{-}+\pi^{+})}$}, for \mbox{A--A} and pp collisions. These ratios are shown in Figure \ref{fig-1}c as a function of $\langle{N_{\rm part}\rangle}$, both at LHC and RHIC energies. The relative production of strangeness in pp collisions at the LHC is larger than at RHIC energy while it is almost the same at the two energies for \mbox{A--A} collisions. The decrease of enhancement with increasing energy is caused by the change in \mbox{pp} collisions, in agreement with the hypothesis of progressive removal of canonical suppression in \mbox{pp} collisions with increasing energy. The hyperon-to-pion ratio increases when going from pp to \mbox{A--A}, showing a relative enhancement of strangeness production in \mbox{A--A} collisions (normalized to the pion yield) which is about half of that seen when normalizing to the mean number of participants. The enhancement rises with centrality up to about $\langle{N_{\rm part}\rangle} \sim{150}$ and apparently saturates thereafter. 

\begin{figure}
\centering
\includegraphics[width=14pc,clip]{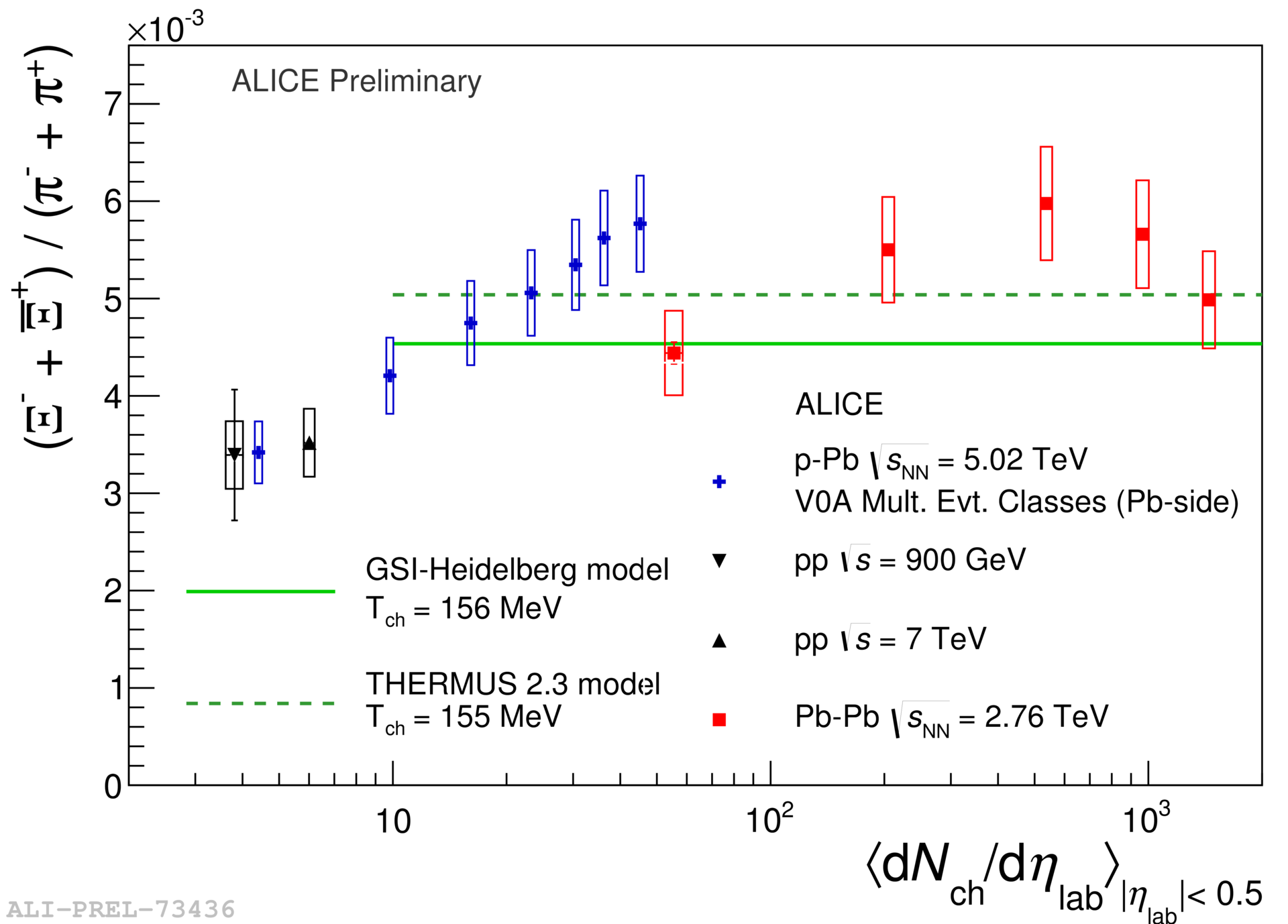}
\includegraphics[width=14pc,clip]{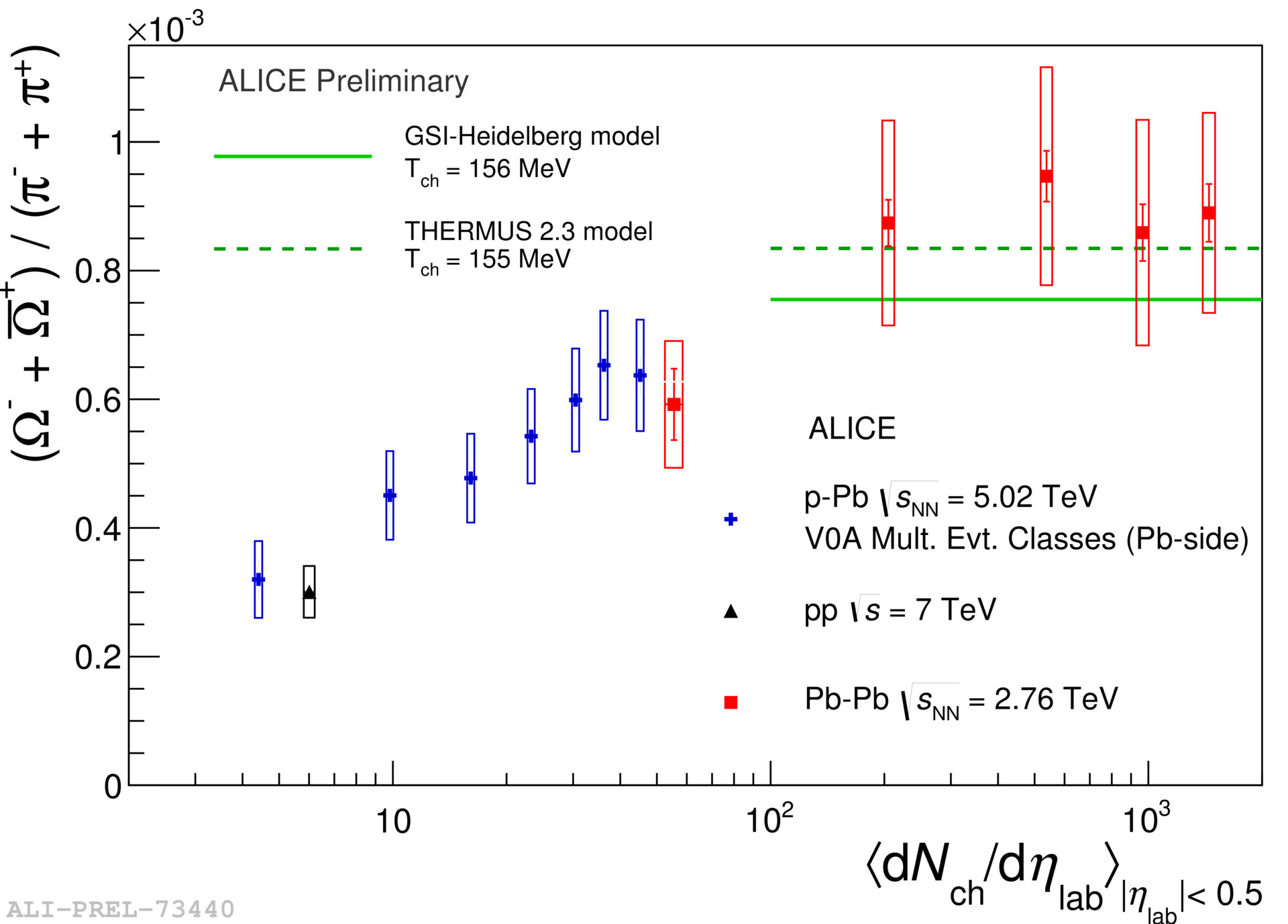}
\caption{$(\Xi^{-}+\overline{\Xi}^{+})/(\pi^{-}+\pi^{+})$ and $(\Omega^{-}+\overline{\Omega}^{+})/(\pi^{-}+\pi^{+})$ ratios as a function of charged-particle pseudorapidity density in \mbox{pp}, \mbox{p--Pb} and \mbox{Pb--Pb} collisions measured by ALICE.}
\label{fig-2}  
\end{figure}

Figure \ref{fig-2} shows the recently measured \mbox{p--Pb} hyperon-to-pion ratios as a function of the charged-particle density, compared with results from \mbox{pp} and \mbox{Pb--Pb} collisions. The multiplicity in \mbox{p--Pb} collisions is intermediate between \mbox{pp} and \mbox{Pb--Pb}. An enhancement of the multi-strange baryons with multiplicity in \mbox{p--Pb} data is observed; it still increases with the strangeness content of the baryon, also considering the earlier 2$\Lambda$/($\pi^{-}$+$\pi^{+}$) ratios published in \cite{lfpPb}.

In Figure \ref{fig-1}c the predictions for the hyperon-to-pion ratios from thermal models, based on a grand canonical approach are also shown: GSI-Heidelberg model described in \cite{GSImodel} (full line, with a chemical freeze-out temperature parameter \mbox{T = 164 MeV}) and THERMUS described in \cite{THERMUS} (dashed line, with \mbox{T = 170 MeV}). Those temperatures were obtained from extrapolations of RHIC results to LHC energies. Recently,  a global fit of all the hadron yields, measured by the ALICE Collaboration, has also been performed \cite{thermalRHICvsLHC} with the thermal grand canonical model; the chemical equilibrium temperature parameter obtained from the simultaneous fit is lower than the one extrapolated from low energy results and lies at \mbox{T = 156 MeV}. The solid (GSI-Heidelberg model) and dashed (THERMUS) green lines in Figure 2 show ratios corresponding to chemical equilibrium temperatures of 156 MeV and 155 MeV, respectively. Except for the most peripheral \mbox{Pb--Pb} data point, the saturation limits tend to overlap with the lower end of the \mbox{Pb--Pb} data points error bars or lie below the measured data. In \mbox{p--Pb} the $\Xi/\pi$ ratios exceed the thermal limits imposed by the global fits, in contrast to the $\Omega/\pi$ ratios which does not reach the equilibrium level as seen for the most peripheral \mbox{Pb--Pb} collisions.

\section{Multi-strange nuclear modification factors}
\label{sec-3}
\begin{figure}[b!]
\centering
\includegraphics[width=14pc,clip]{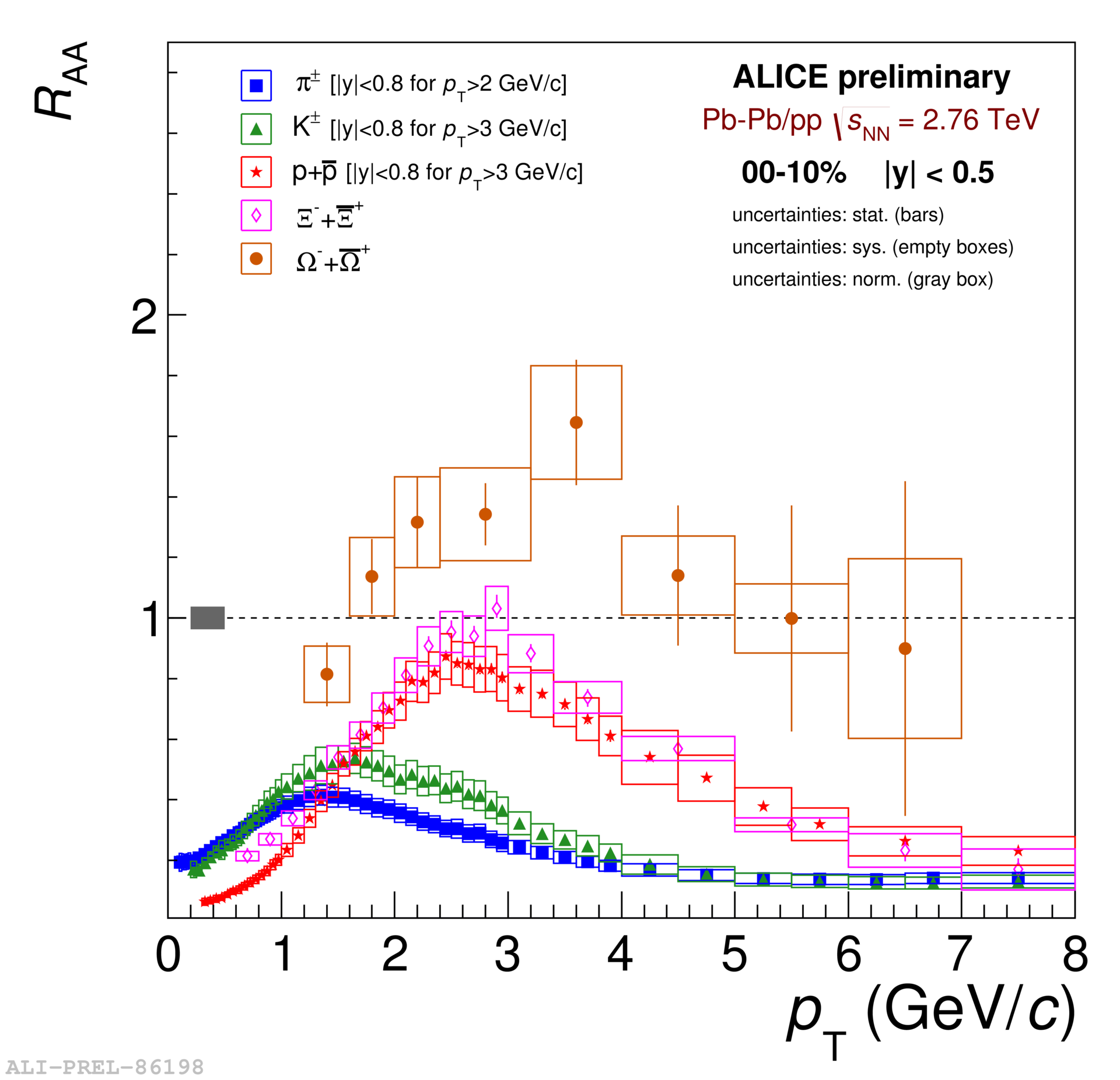}
\includegraphics[width=14pc,clip]{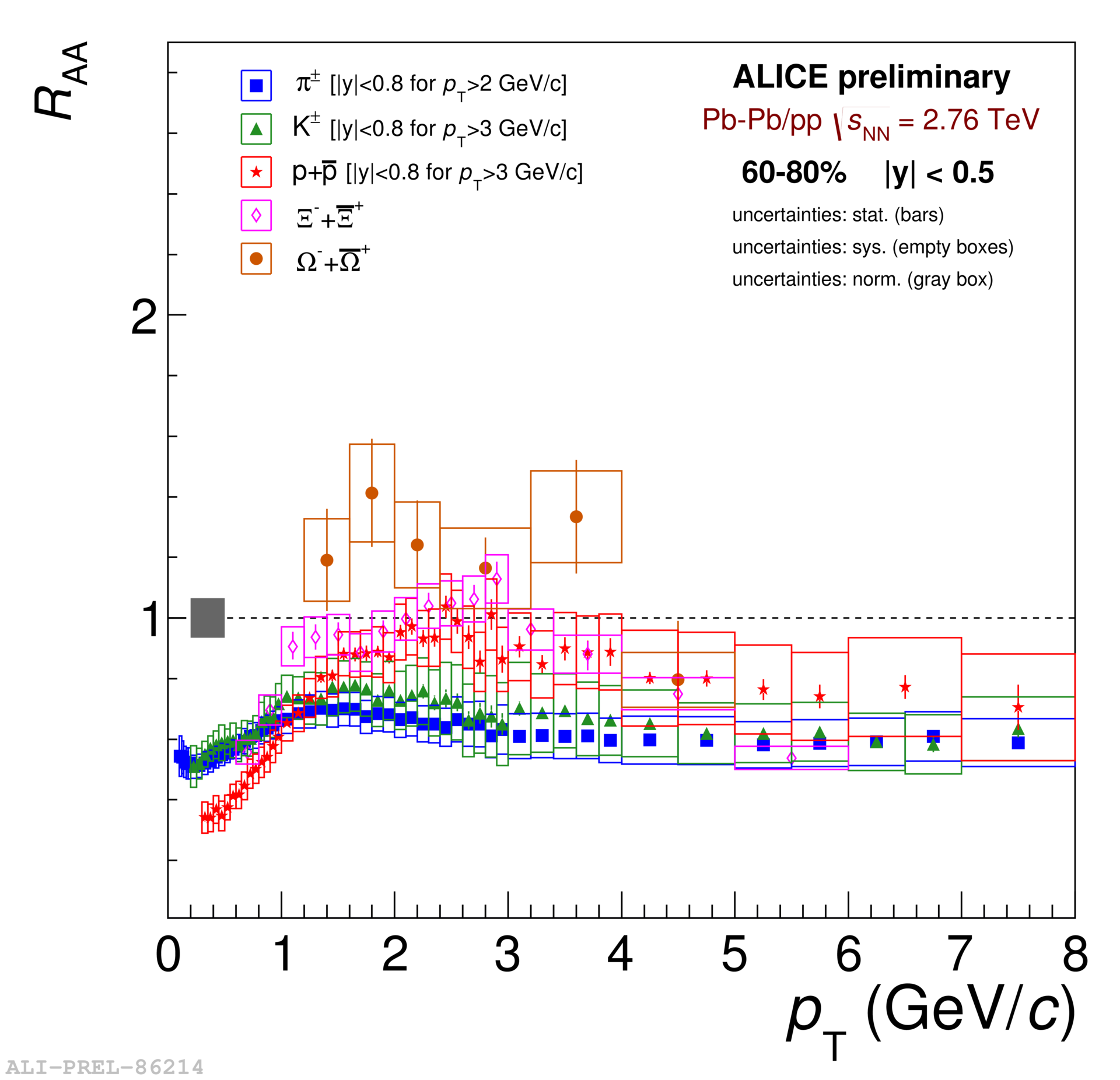}
\caption{Nuclear modification factors for $\Xi$ and $\Omega$, compared to the ones for $\pi$, K and p, as a function of $p_{\rm T}$ in 0-10\% (left) and 60-80\% (right) centrality \mbox{Pb--Pb} collisions at $\sqrt{s_{NN}}$ = \mbox{2.76 TeV}.}
\label{fig-3}  
\end{figure}

The nuclear modification factor ($R_{\rm AA}$) is defined as the ratio of the $p_{\rm T}$ spectra in \mbox{Pb--Pb}  and in \mbox{pp} collisions scaled by the number of nucleon-nucleon collisions \cite{centrality}. In Figure \ref{fig-3} the $R_{\rm AA}$ as a function of the $p_{\rm T}$ for multi-strange baryons, in the most central (0-10\%) and most peripheral (60-80\%) collisions, are shown and compared with lighter hadrons ($\pi$, K and p). 

The spectra have been measured in \mbox{|\textit{y}| < 0.5} for \mbox{$p_{\rm T}$ < \mbox{2 GeV/\textit{c}}} and \mbox{3 GeV/\textit{c}} respectively in case of $\pi$ and K, p; above these values the rapidity window \mbox{|\textit{y}| < 0.8} has been used. The spectra for the cascades have been all measured in \mbox{|\textit{y}| < 0.5}. Moreover, in the most central class, the last two points of the $R_{\rm AA}$, both for the $\Xi$ and the $\Omega$, have been obtained extrapolating the reference \mbox{pp} spectra using a L\'evy-Tsallis fit function. 

In the most central class (left panel), the $R_{\rm AA}$ for $\Xi$ follows the same trend as the proton at high $p_{\rm T}$ ($>$ 6 GeV/\textit{c}), where the suppression does not depend on the particle mass. At intermediate $p_{\rm T}$ there are indications of mass-ordering among the baryons, that can be interpreted as consequence of the radial flow \cite{Raa}. The $R_{\rm AA}$ for the $\Omega$ is above unity, which might be the results of the larger contribution of the strangeness enhancement compared to  $\Xi$. 
When going from the most central to the most peripheral class (right panel) the $R_{\rm AA}$ values and the relative difference between the particles are reduced. As expected, also the strangeness enhancement effect for the $\Omega$ becomes weaker.

\section{Conclusions}
\label{sec-4}
The production of the multi-strange baryons $\Xi^{-}$ and $\Omega^{-}$ and their antiparticles have been measured by the ALICE Collaboration in \mbox{pp}, \mbox{p--Pb} and \mbox{Pb--Pb} collisions. 
A strangeness enhancement has been observed in the \mbox{Pb--Pb} collisions with characteristics similar to the ones already observed at lower energies; the decreasing trend with the increasing of the centre-of-mass energy has been confirmed. The hyperon-to-pion ratios measured in \mbox{p--Pb} collisions bridge the values in \mbox{pp} to those in \mbox{Pb--Pb} and reveal a dependency of the strangeness enhancement on the charged particle multiplicity. The comparison to the grand canonical thermal model expectations in \mbox{Pb--Pb} collisions for \mbox{T = 155 MeV}, gives, within a general agreement with the measurements in \mbox{Pb--Pb} collisions, a tension between ratios for $\Xi$ and $\Omega$ in \mbox{p--Pb} collisions: while $\Omega/\pi$ only approaches the saturation level with increasing multiplicity, the $\Xi/\pi$ values at high multiplicity are comparable to the \mbox{Pb--Pb} values and lie above the equilibrium limits. 
Finally, the $R_{\rm AA}$ shows a large suppression with similar values for all the species at \mbox{$p_{\rm T} >$ 6 GeV/\textit{c}}, presence of mass ordering at intermediate $p_{\rm T}$ and a clear strangeness enhancement effect especially for the $\Omega$ .



\begin{thebibliography}{}

\bibitem{Rafelski} J.~Rafelski, B.~M\"uller, \emph{Phys. Rev. Lett.} \textbf{48} 1066 (1982). \\
P.~Koch, J.~Rafelski, W.~Greiner, \emph{Phys. Lett. B} \textbf{123} 151 (1983). \\
P.~Koch, B.~M\"uller, J.~Rafelski, \emph{Phys. Rep.} \textbf{142} 167 (1986).
\bibitem{strangEnhancNA57}
NA57 Collaboration F. Antinori {\it et al.}, \emph{J. Phys. G} \textbf{32} 427 (2006). \\ 
NA57 Collaboration F. Antinori {\it et al.}, \emph{J. Phys. G} \textbf{37} 045105 (2010). 
\bibitem{strangEnhancSTAR}
STAR Collaboration B. I. Abelev {\it et al.},  \emph{Phys. Rev. C} \textbf{77} 044908 (2008). 
\bibitem{strangEnhancALICE}
ALICE Collaboration B.~Abelev {\it et al.}, \emph{Phys. Lett. B} \textbf{728} 216 (2014).  \\
ALICE Collaboration B.~Abelev {\it et al.}, \emph{Phys. Lett. B} \textbf{734} 409 (2014). 
\bibitem{RHICRaa}
PHENIX Collaboration K.~Adcox {\it et al.}, \emph{Phys. Rev. Lett.} \textbf{88} 022301 (2002). \\
STAR Collaboration C.~Adler {\it et al.}, \emph{Phys. Rev. Lett.} \textbf{89} 202301 (2002).
\bibitem{ALICERaa}
ALICE Collaboration K.~Aamodt {\it et al.}, \emph{Phys. Lett. B} \textbf{696} 30 (2011). 
\bibitem{JINST}
ALICE Collaboration K.~Aamodt {\it et al.}, \emph{JINST} \textbf{3} S08002 (2008).
\bibitem{lfpPb}
ALICE Collaboration B.~Abelev {\it et al.}, \emph{Phys. Lett. B} \textbf{728} 25-38 (2014).
\bibitem{GSImodel}
A. Andronic, P. Braun-Munzinger, J. Stachel, \emph{Phys. Lett. B} \textbf{673} 142 (2009); \\
A. Andronic, P. Braun-Munzinger, J. Stachel, \emph{Phys. Lett. B} \textbf{678} 516 (2009) (Erratum).
\bibitem{THERMUS}
J. Cleymans, I. Kraus, H. Oeschler, K. Redlich, S. Wheaton, \emph{Phys. Rev. C} \textbf{74} 034903 (2006).
\bibitem{thermalRHICvsLHC}
J. Stachel, A Andronic, P. Braun-Munzinger, K Redlich, \emph{arXiv:1311.4662v1}.
\bibitem{centrality}
ALICE Collaboration B. Abelev {\it et al.}, \emph{Phys. Rev. C} \textbf{88} 044909 (2013).
\bibitem{Raa}
ALICE Collaboration B. Abelev {\it et al.}, \emph{arXiv:}1401.1250.

\end{thebibliography}
\end{document}